\documentclass[%
 reprint,
 amsmath,amssymb,
 aps,soul,
pre,
]{revtex4-2}

\usepackage{graphicx}
\usepackage{dcolumn}
\usepackage{bm}
\usepackage{xcolor}
\usepackage{booktabs}

\renewcommand{\selectlanguage}[1]{}

\begin{document}

\preprint{APS/123-QED}

\title{Meter-scale Observations of Equatorial Plasma Turbulence}

\author{Magnus F Ivarsen}
\altaffiliation[Also at ]{
The European Space Agency Centre for Earth Observation, Frascati, Italy}
\affiliation{Department of Physics and Engineering Physics, University of Saskatchewan, Saskatoon, Canada}%

\author{Lasse BN Clausen}
\author{Yaqi Jin}
\affiliation{Department of Physics, University of Oslo, Oslo, Norway}%




\author{Jaeheung Park}
\altaffiliation[Also at ]{Department of Astronomy and Space Science, Korea University of Science and Technology, Daejeon, South-Korea}
\affiliation{Space Science Division, Korea Astronomy and Space Science Institute, Daejeon, South-Korea}%

\date{\today}

\begin{abstract}
The multi-Needle Langmuir Probe collects an electron current through four fixed-bias cylindrical copper needles. This allows for an extremely high sampling frequency, with plasma properties being inferred through polynomial fitting in the current-voltage plane. We present initial results from such a multi-needle probe mounted on the International Space Station, orbiting Earth at an altitude of around 400~km. That altitude, and its orbital inclination ($\sim50^\circ$), place the ISS as a suitable platform for observing equatorial plasma bubbles. In case studies of such turbulent structuring of the F-region plasma, we observe density timeseries that conserve considerable detail at virtually every level of magnification down to its Nyquist scale of 2--5~meters. We present power spectral density estimates of the turbulent structuring found inside equatorial plasma bubbles, and we discuss apparent break-points at scale-sizes between 1~m and 300~m, which we interpret in the light of turbulent dissipation as kilometer-scale swirls produced by the gradient-drift instability dissipate in the plasma.
\end{abstract}

\maketitle


\section{\label{sec:intro}Introduction}

In space plasmas, turbulence manifests as a complex, multi-scale interaction between various chaotic fluctuations in electromagnetic fields and the ionized gas density \cite{tsunodaHighlatitudeRegionIrregularities1988,hubaIonosphericTurbulenceInterchange1985}. This daunting description can however be made considerably simpler by describing plasma turbulence as an irregular \textit{current}, or river, of charged particles breaking-up into smaller and smaller swirls, with the initial kinetic energy of the irregular flow eventually transitioning into heat, as the charged particles collide with the much-more numerous neutral gas particles. 

In an idealized sense, this principle of breaking-up the otherwise laminar flow is encapsulated by the term `turbulent cascade', wherein every spatial scale (the size of the swirl) transfers an amount of energy to its \textit{adjacent} spatial scales equal to the energy it in turn received. Or, in the words of Fourier analysis, the power spectral density (PSD) decreases in a monotonous way with increasing \textit{wavenumber}. Consequently, the decay in PSD of a stable, turbulent fluid process is proportional to the wavenumber raised to the power -5/3, the ``Kolmogorov five-thirds'' \cite{kolmogorov_local_1968}. Although such behaviour can be seen in isolated cases, the real state of the turbulent ionosphere is in general a complicated non-linear state of both injection and dissipation of free energy    \cite{chaturvediNonlinearStabilizationCurrent1979,ivarsen_what_2024}.

\begin{figure}[b]
    \centering
    \includegraphics[width=0.5\textwidth]{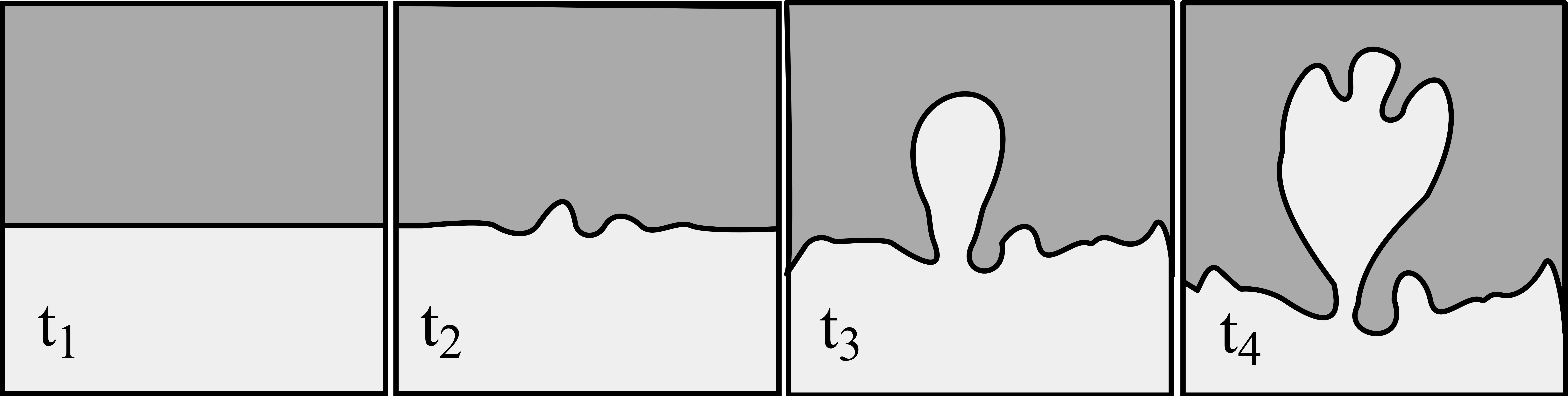}
    \caption{Kelley's cartoon of the creation and updrift of an equatorial plasma bubble, shaped by the Rayleigh-Taylor instability. The bubbles initially channel through the thick F-region ionosphere,  ``blossoming'' on the topside, where various instabilities cause turbulent structuring on the walls of the bubbles. Drawn after Ref.~\cite{kelley_chapter_1989}.}
    \label{fig:cartoon}
\end{figure}

The most turbulent part of geospace is arguably the equatorial ionosphere as it appears after sunset \cite{woodmanRadarObservationsRegion1976,heelisElectrodynamicsLowMiddle2004,kilMorphologyEquatorialPlasma2015}. The Rayleigh-Taylor instability seeds wide, field-aligned holes, or bubbles, in the plasma density that rise upwards by merit of having a lower density than the surrounding plasma \cite{kilEquatorialDensityIrregularity1998}, a process that is illustrated in Figure~\ref{fig:cartoon}. The action is known to trigger a host of attendant irregularities in the plasma along the wall of the bubble \cite{hysellCollisionalShearInstability2004,tulasi_ram_dilatory_2020}. This process is turbulent in the proper sense of the word and ensures that the shape of such plasma bubbles feature distinct, though similar, structure on a wide range of spatial scales \cite{ivarsen_plasma_2024}.

As alluded to, turbulent structures in the ionosphere are readily scrutinized with PSD analysis, wherein a signal is decomposed into its Fourier constituents. The turbulent cascade, the decrease in power with increasing wavenumber, can then be expressed as,
\begin{equation} \label{eq:slope}
P(k) \propto  k^{-n},
\end{equation}
with a wavenumber $k$, its PSD $P(k)$, and the spectral index denoted by $n$, a real-valued exponent. $n$ is frequently the topic of studies that seek to understand equatorial plasma irregularities \cite{dysonSituMeasurementsSpectral1974,rinoSimultaneousRocketborneBeacon1981,kelleyEquatorialSpreadFReview1981,kilEquatorialDensityIrregularity1998}. A spectral break occurs when the PSD is more accurately described by a composite powerlaw,
\begin{equation} \label{eq:dualslope}
P(k) \propto 
\begin{cases}
 k^{-n_1}, & \text{when } k\leq k_0, \\
 k^{-n_2}, & \text{when } k> k_0,
\end{cases}
\end{equation}
with $k_0$ acting as a break-point. The wavenumber $k$ is related to scale-size $\lambda$ through,
\begin{equation}
    k = \frac{2\pi}{\lambda}.
\end{equation}

In the literature, spectral breakpoints in equatorial plasma bubbles have been identified at various scales between 1 km and 1 m \cite{rinoSimultaneousRocketborneBeacon1981,kelleySimultaneousRocketProbe1982,basuHighResolutionTopside1983,labelleAnalysisRoleDrift1986,hysellSteepenedStructuresEquatorial1994,kilEquatorialDensityIrregularity1998,suROCSATIonosphericPlasma2001,lePlasmaDensityEnhancements2003,rinoWaveletbasedAnalysisPower2014,rinoCharacterizationIntermediatescaleSpread2016}. Crucially, such spectral breaks can be considered as energy injections \cite{zalesakNonlinearEquatorialSpread1982,labelleGenerationKilometerScale1986,kilEquatorialDensityIrregularity1998}, in which PSD above the break point is lifted nominally from the otherwise steep decay in power that signifies dissipation \cite{kivancSpatialDistributionIonospheric1998,keskinenNonlinearEvolutionHighlatitude1990,ivarsenDirectEvidenceDissipation2019}. A small selection of characteristic, spectral breakpoint observations in equatorial plasma bubbles are summarized in Table~\ref{table:studies}.

\begin{table}[t!]
\centering
\begin{tabular}{@{}l|c|c|c|c@{}}
Study & Spectral break & Altitude & Year & Local time \\ \midrule
\cite{rinoSimultaneousRocketborneBeacon1981} & 500 m & $\sim$300 km & 1979 & Near 24 LT \\
\cite{hysellSteepenedStructuresEquatorial1994} & 80 m - 100 m & $\sim$500 km & 1990 & 21 LT \\
\cite{labelleAnalysisRoleDrift1986} & 60 m - 260 m & $\sim$400 km & 1983 & 21:40 LT \\
\cite{suROCSATIonosphericPlasma2001,lePlasmaDensityEnhancements2003} & \textless 100 m & 600 km & 1999 - 2000 & 20 LT -22 LT \\
\cite{rinoWaveletbasedAnalysisPower2014,rinoCharacterizationIntermediatescaleSpread2016} & 100 m - 10 km & $<$800 km & 2011 - 2014 & Pre-24 LT \\ \midrule
\end{tabular}
\caption{A summary of spectral breaks found in five studies of equatorial plasma bubbles. \label{table:studies}}
\end{table}

The observations of breakpoints on scale-sizes smaller than around $\sim300$~m are important for the observed distribution of satellite communication signal scintillations (stochastic variations in the amplitude of the pseudo-random signal emitted by navigational system satellites, frequently referred to as `GPS scintillations') \cite{kintnerp.m.GPSIonosphericScintillations2007,mezianeTurbulenceSignaturesHighLatitude2023}, a characteristic radio signature of the sunset equatorial ionosphere \cite{basuEquatorialScintillationsAdvances1985,tsunodaControlSeasonalLongitudinal1985,yehRadioWaveScintillations1982}, and one that poses societal, technological challenges for the future \cite{moenSpaceWeatherChallenges2013}. In contrast, a fully steepened ($n\gg5/3$) spectrum around the $\sim300$~m scale-size is indicative of turbulence that is being (fully or partially) dissipated into heat \cite{st-mauriceSmallScaleIrregularities2009,st-mauriceRevisitingBehaviorERegion2021}. 

\begin{figure}
    \centering
    \includegraphics[width=0.5\textwidth]{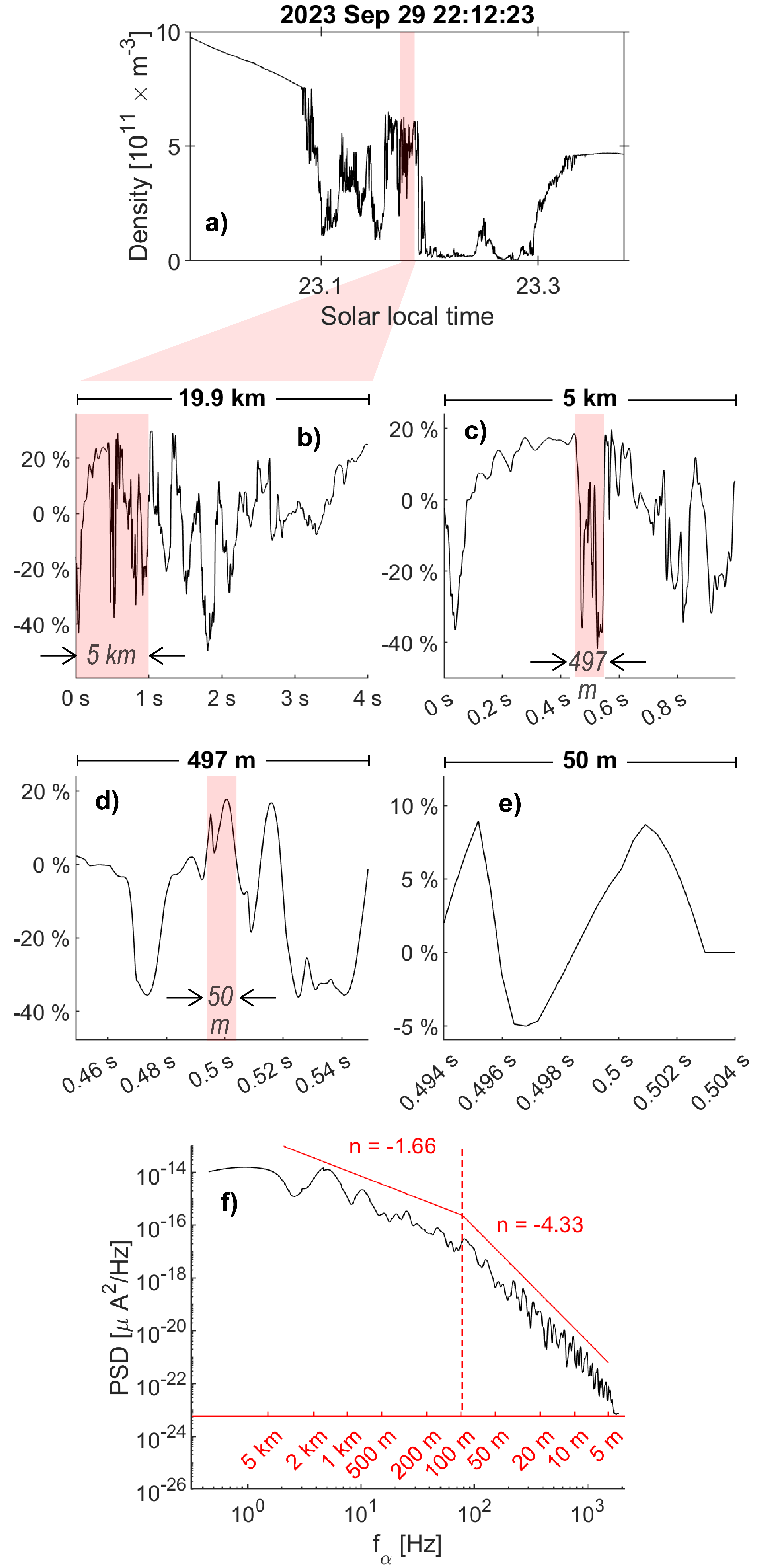}
    \caption{Panel a) shows a 2.5~kHz density timeseries observed by the m-NLP instrument onboard the ISS. Panels b--e) consist of successively enlarged sub-intervals of the full timeseries,  expressed in terms of \% deviation from the median value inside the frame. Panel f) shows a PSD analysis, with scale-sizes indicated with an additional, red $x$-axis, adjusted for the angle between the space station's velocity vector and the local magnetic field vector \cite{ivarsen_plasma_2024}. Break-points and spectral indices are indicated (see Eqs.~\ref{eq:slope} -- \ref{eq:dualslope}).}    
    \label{fig:case1}
\end{figure}


This paper presents an analysis of case studies of equatorial plasma bubbles, based on super-high resolution ($\geq2.5$~kHz) \emph{in-situ} fluctuations in plasma density, performed by the multi-Needle Langmuir Probe (m-NLP) instrument newly mounted on the International Space Station (ISS). Although observations have previously been made with comparable resolution, such observations have hitherto largely been presented in technical papers (e.g., \cite{pfaff_vector_2021}). We discuss the occurrence of spectral breakpoints in equatorial plasma turbulence, where a tendency for spectra to break at scale-sizes between 50~m--300~m point towards the gradient-drift instability as being instrumental in producing the observed, post-sunset equatorial ionosphere.

\section{Methods}

We analyze the current collected by four needle Langmuir probes organized in an m-NLP instrument \cite{hoangMultiNeedleLangmuirProbe2018,liuMNLPInferenceModels2023}. A different, fixed bias is applied to all four probes, meaning that plasma properties can be inferred by extrapolating the true current-voltage curve obtained by a sweeping Langmuir probe based on four current-voltage measurements (see Figure~3 in \cite{ivarsenMultineedleLangmuirProbe2019}). As alluded to, the instrument has newly been mounted on the ISS, where it is capable of measuring plasma density and temperature with a measurement frequency of up to 10~kHz. In the present letter, we analyze the collected current with a measurement frequency of 2.5~kHz.

Figure~\ref{fig:case1}a) shows an example observation of the plasma density in and around equatorial plasma bubbles, for a two-minute stretch of orbit on 29 September 2023, prior to local midnight, at a magnetic latitude around 20$^\circ$ over South-America. We analyze such timeseries under the assumption that the space station velocity (7.66~km/s) far exceeds the local plasma drift velocity ($<1$~km/s \cite{woodmanEastwestIonosphericDrifts1972}).  Then, we calculate the 3D-angle between spacecraft displacement and the local magnetic field vector to derive the field-perpendicular velocity of the space station. This angle scales the effective field-perpendicular scale-sizes available to scrutiny; see Eqs.~(2--4) in Ref.~\cite{ivarsen_plasma_2024}. 

Next, we calculate the relative density fluctuations,
\begin{equation} \label{eq:I}
    \Delta I(t) = \frac{I(t)}{I_\tau(t)} - 1,
\end{equation}
where $I(t)$ is the collected current of ions at time $t$ and $I_{\tau}(t)$ is the median value of $I(t)$ in a window of length $\tau$ centered on time $t$; $\Delta I(t)$ then represents local density fluctuations. We plot the current fluctuations in successively enlarged segments, where each successive segment is divided by a decreasing window length $\tau$, and displayed in a plot window of length $\tau$ (Figure~\ref{fig:case1}b--e). Next, to isolate the contributions of turbulent swirls of various sizes in producing these timeseries, we also apply a PSD analysis to the $I(t)$ timeseries, using a variant of overlapping, averaged periodograms (Welch's method) \cite{trobsImprovedSpectrumEstimation2006}, following Refs.~\cite{ivarsenDirectEvidenceDissipation2019,ivarsenLifetimesPlasmaStructures2021,ivarsen_plasma_2024}. We identify the spectral slopes and break-points (Eq.~\ref{eq:dualslope}) by minimizing the root-mean-square error of piece-wise linear fits \cite{ivarsen_what_2024}. An example of this method is on display in Figure~\ref{fig:case1}f), showing the PSD of the 4~second timeseries in panel b). A spectral break-point is identified at around 100~m; the decay in power for scales larger than this break-point is consistent with a turbulent cascade ($n\approx5/3$). Thus, the self-similar sawtooth patterns that are repeating in Figure~\ref{fig:case1}b--e) is a technically correct illustration of the turbulent cascade.

\begin{figure}[h!]
    \centering
    \includegraphics[width=.5\textwidth]{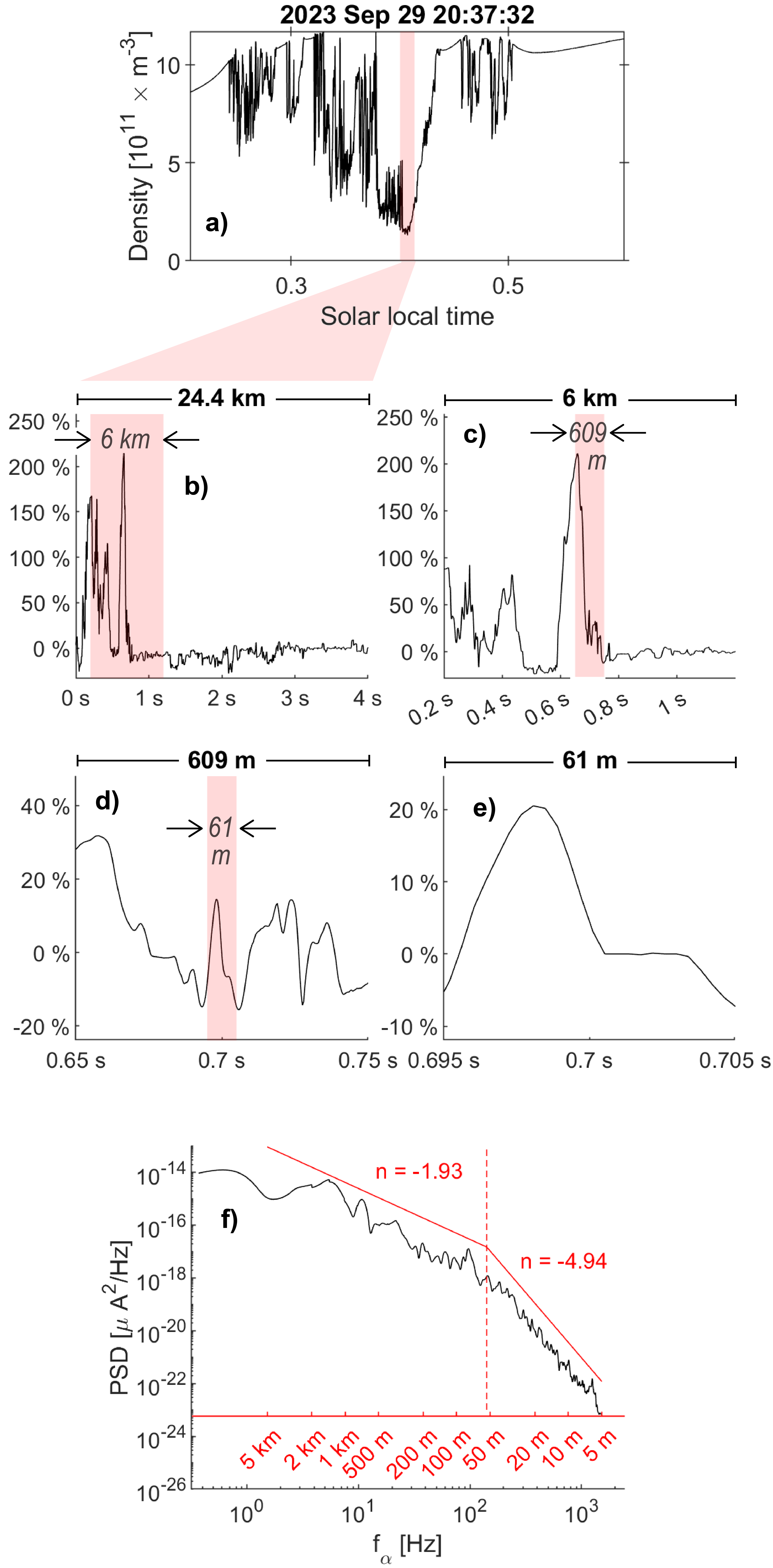}
    \caption{Observations of equatorial plasma bubbles (panel a), the fractal patterns of the fluctuations inside a 4~s interval (panels b--e), and a PSD analysis of the 4~s interval (panel f).}    
    \label{fig:case2}
\end{figure}

\section{Results}

Having introduced the measurements and the analysis techniques we shall now present two additional case studies that combine the fractal properties discussed above with so-called energy injections, small-scale fluctuations in the timeseries that cannot be explained in terms of a turbulent cascade.

Figure~\ref{fig:case2} shows, in a mode of representation entirely equivalent to Figure~\ref{fig:case1}, the current fluctuations measured inside and around plasma bubbles observed earlier in the evening of 29 September 2023, just after local midnight at a magnetic latitude of around 15$^\circ$ over the Indian Ocean. We observe that a spike-like feature is \textit{repeated at various magnifications} between 24.4~km and 61~m (Figure~\ref{fig:case2}b--e), a strong indication that the break-point at around 50~m in Figure~\ref{fig:case2}f) is indicative of an energy injection occurring at some higher scale, perhaps several kilometers of size. Inspecting both Figure~\ref{fig:case1}a) and Figure~\ref{fig:case2}a), we note the appearance of sharp, kilometer-size spikes on the edges of the density depressions in both cases.


\begin{figure}
    \centering
    \includegraphics[width=0.5\textwidth    ]{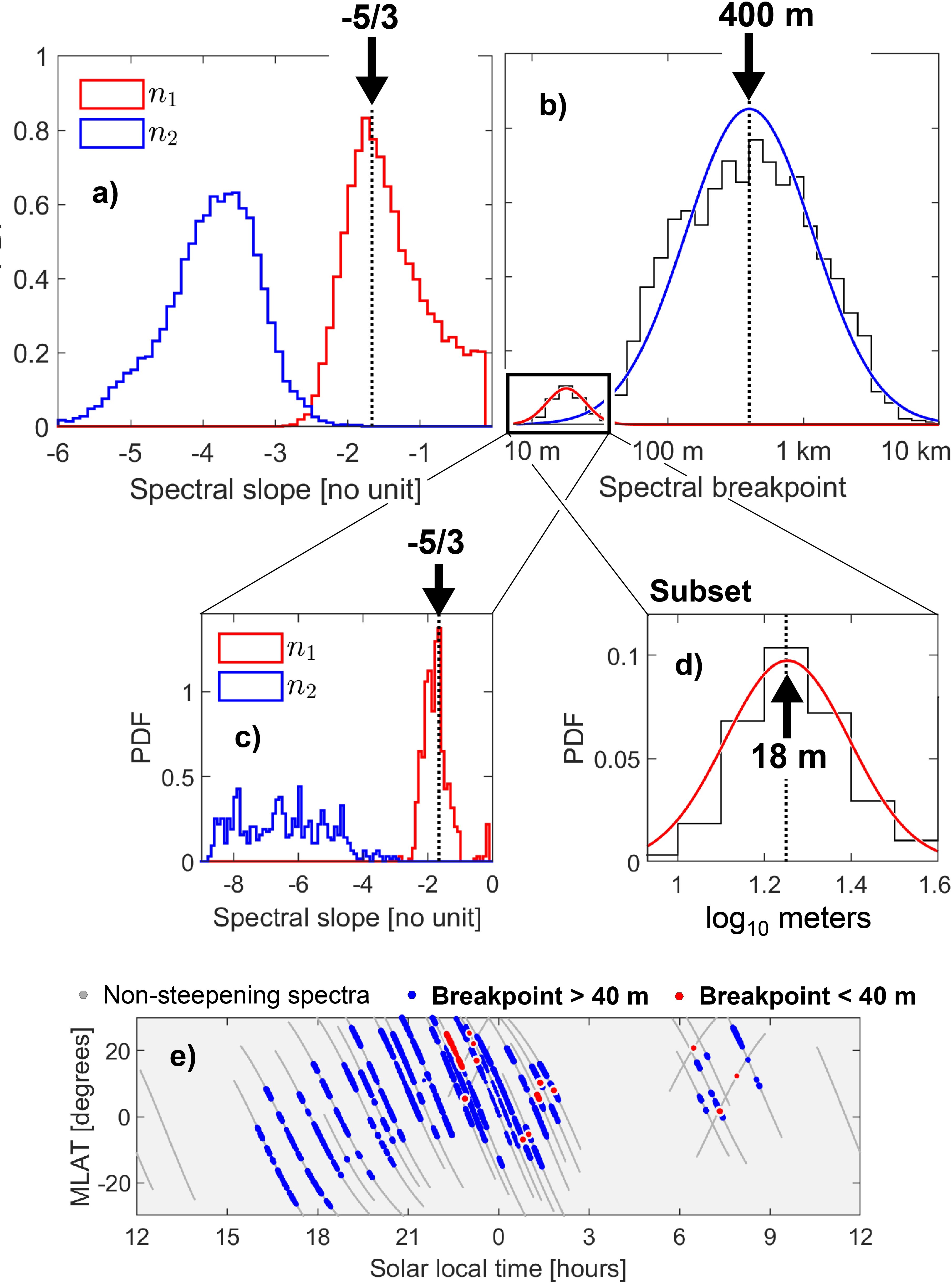}
    \caption{Histograms of steepening density spectra, following Ref.~\cite{ivarsenSteepeningPlasmaDensity2021} in defining `spectral steepening' as having a difference in spectral indices greater than 0.9 in the concave direction. Panel a) compare the first and second spectral index, while panel b) shows the spectral breakpoint locations. The statistics is based on some 20,000 spectra measured between 29 September and 4 October 2023. Panels c--d) show equivalent statistics for the subset of spectra that exhibit a breakpoint smaller than 40~m. Panel e) displays where the various spectra were measured, in a solar local time-magnetic latitude coordinate system, representing the subset of spectra that exhibit small-scale breakpoints with red color.}
    \label{fig:stats}
\end{figure}

Figure~\ref{fig:stats} show preliminary statistics, based on six days of operation during a particularly active period of Fall 2023, we have collected some 20,000 spectra that we deem as stemming from equatorial plasma turbulence, following the criteria in Ref.~\cite{ivarsen_plasma_2024}. Figure~\ref{fig:stats}a) shows histograms of steepening density spectra (Eq.~\ref{eq:dualslope}), with $n_1$ in red and $n_2$ in black. The ``Kolmogorov -5/3'' is marked with a dotted, black line, and we note that $n_1$ is distributed in a remarkably sharp fashion around that value -- though situated slightly short of -5/3 on the negative side. Panel b) shows the spectral breakpoint locations (converted to spatial scale), with the prominent peak of 400~m marked by dotted lines. A secondary peak is evident at 18~m. In red and blue, we show normal distributions for the small- and large-scale peaks respectively. For the prominent, wide peak around 400~m, we refer back to Table~\ref{table:studies} and note the excellent agreement between our results the literature of irregularity spectral steepening in equatorial plasma observations. The smaller-scale distribution, however, is less anticipated by the scientific body of literature. Panels c--d) show equivalent statistics for the subset having spectral breakpoints smaller than 40~m, noting that for the $\sim3$~\% of our spectra that exhibit a small-scale injection point, the initial slope $n_1$ is sharply distributed around -5/3. Figure~\ref{fig:stats}e) displays the solar local times (at measurement) of the various spectra, showing that the subset of spectra that feature small-scale breakpoints are observed around and after midnight.



\begin{figure}
    \centering
    \includegraphics[width=.485\textwidth]{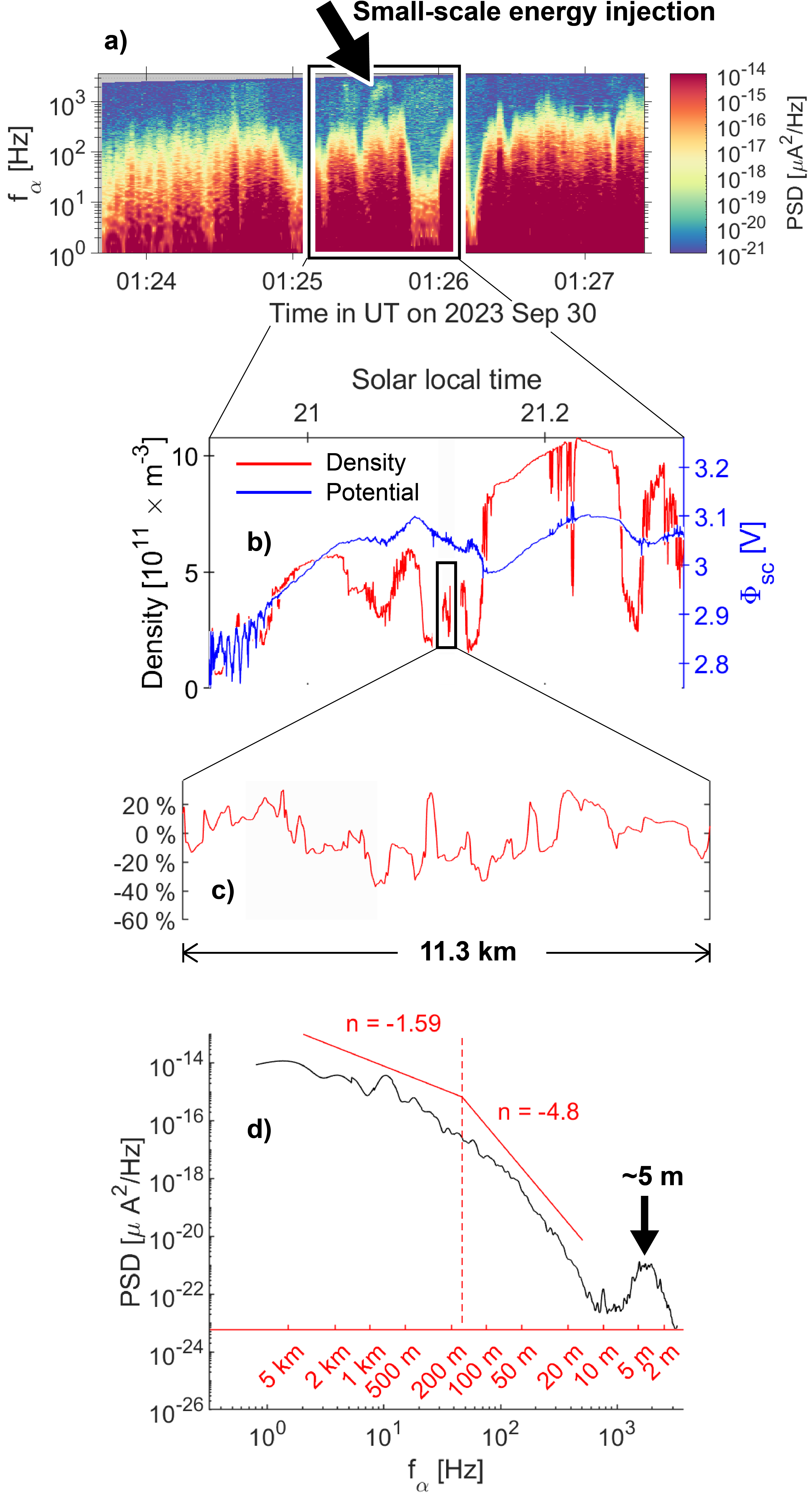}
    \caption{Observations of equatorial plasma bubbles, now showing a spectrogram of power per frequency bin ($y$-axis) and per 0.5~second timebin ($x$-axis), with a logarithmic colorscale. Panel b) shows density in red and spacecraft potential in blue, for a $\sim1$~minute interval of panel a). Panel c) shows density inside a \textit{four-second} interval from the former panel (11.3 km in field-perpendicular distance). Panel d) shows the calculated powerspectrum of the ion current fluctuations during the four-second interval, with spectral index information posted. A prominent ``bump'', or energy injection, is apparent at around 5~m scale-size.}    
    \vspace{-20pt}
    \label{fig:case3}
\end{figure}

We shall next analyze that subset of our observations, the spectra containing a small-scale injection point, noting that it is a more ambiguous, but potentially more important finding the for future developments in the study of equatorial plasma turbulence. Figure~\ref{fig:case3}, a typical example taken from that subset, shows a plasma density timeseries along a short stretch of ISS orbit. Panel a) displays a spectrogram of all the powerspectra observed during a $\sim$4~minute stretch of orbit, with a colorscale chosen to highlight any small-scale features. The next three panels hones in on a particular, localized small-scale feature of the spectrogram. The four-second segment (Figure~\ref{fig:case3}c), and its attendant power spectral density estimate (Figure~\ref{fig:case3}d), exhibit a conspicuous ``bump'' of elevated power for a narrow range of wavenumbers (spatial scale around $\sim5$~m). The appearance of such bumps are indicative of strong energy injections or the characteristic trigger of a specific instability. In Figure~\ref{fig:case3}, we observe a bump at the frequency corresponding to 5~meters.

\section{Discussion}

Observations of such small-scale fluctuations in equatorial plasma density are almost absent in the literature, due to the general lack of super-high resolution measurements from orbiting platforms ($>2.5$~kHz) and scarcity of rocket experiments with such high-resolution measurements. As a result, the observations of meter-scale energy injections in our database of observations from the ISS are difficult to interpret. 
The subset is small; the small-scale bumps appear intermittently throughout each of the days under study, appearing in around 3\% of the analyzed spectra.

Hence, we ascertain with some confidence that the subset of observations that we have found to exhibit a meter-scale bump in irregularity power represent unprecedented observations of instability-driven meter-scale turbulence on the edges of equatorial plasma bubbles. A candidate for such structuring in the F-region is turbulence triggered by small-scale instabilities secondary to the Rayley-Taylor and gradient-drift instabilties \cite{tulasi_ram_dilatory_2020}; akin to equatorial electrojet turbulence at lower altitudes \cite{oppenheimEvidenceEffectsWavedriven1997}. Lastly, we cannot rule out the possibility of kinetic instabilities. Should future studies substantiate our findings, the ISS m-NLP dataset can be of substantial utility for future attempts at modeling small-scale turbulence in the equatorial region, as small-scale ($<100$~m) turbulence is difficult to resolve with current modeling efforts \cite{yokoyama_review_2017}.

At any rate, the clear distribution of break-point around 400~m evident in Figure~\ref{fig:stats}b) is a strong indication that the gradient-drift instability (GDI) is routinely being triggered and subsequently injecting kilometer-scale swirls of plasma into the wall of the bubbles, as the GDI instability in general favours those scales \cite{tsunodaHighlatitudeRegionIrregularities1988}. The spectra are of unusual high quality, and so the ISS m-NLP dataset is of interest to studies into Fresnel-scale irregularities capable of causing GPS amplitude scintillations, as instabilities triggering the formation of swirls with characteristic sizes between 100~m and 1~km are of special interest for such applications \cite{xiong_occurrence_2020,hamzaTwocomponentPhaseScintillation2023,song_investigating_2025}. 

\vspace{-14pt}

\section{Conclusion}

\vspace{-7pt}

The multi-Needle Langmuir Probe (m-NLP) system mounted on the International Space Station is capable of super high-resolution ($>2.5$~kHz) measurements of equatorial plasma turbulence in Earth's ionosphere. The space station's orbital inclination and its altitude in the ionosphere's F-region make it an ideal platform for the observation of the intense plasma turbulence found inside equatorial plasma bubbles. Such bubbles rise up from the bottomside ionosphere during sunset in the equatorial region of Earth, after which the bubbles slowly crumble, creating smaller and smaller swirls of turbulent plasma in the process. The action is known to disturb signal communication between ground- and space-based devices.

We present extensive observations of a variety of such plasma bubbles observed during a short (6~day) period near the Autumn Equinox of 2023. In case studies, we find clear, fractal structuring of the plasma on scales down to around 50~m, an  indication that kilometer-size gradient-drift waves are structuring the plasma, with the turbulent cascade in control over the smaller scale-sizes, until rapid dissipation takes over for scale-sizes lower than around 50~m.

For these lowest scales (10~m--50~m), we find intriguing evidence of a small-scale energy injection. This subset is not extensive (constituting around 3~\% of the steepening spectra), but nevertheless exhibit a tantatively characteristic tendency to appear around and after midnight in the equatorial irregularities we sampled.

The results are not to be considered general-statistical or climatological, rather they are a statistical treatment of an extensive dataset collected during six days with appreciable geomagnetic activity near the Equinox of 2023. In future, conjunction studies between the ISS and ground-based GPS receivers will allow for systematic investigations into how the various small-scale spectral features compare to radio scintillations.

\section*{Acknowledgements}
This work was supported in part by the European Space Agency’s Living Planet grant no 1000012348. We acknowledge the support of the Canadian Space Agency (CSA) [20SUGOICEB], the Canada Foundation for Innovation (CFI) John R. Evans Leaders Fund [32117], the Natural Science and Engineering Research Council (NSERC), the International Space Mission Training Program supported by the Collaborative Research and Training Experience (CREATE) [479771-2016], the Discovery grants program [RGPIN-2019-19135]; and the Digital Research Alliance of Canada [RRG-FT2109].


%

\end{document}